\documentclass[prl,aps,twocolumn,superscriptaddress,floatfix]{revtex4-1}

\usepackage{amssymb}
\usepackage{epsfig}
\usepackage{amsmath}
\usepackage{subfigure}
\usepackage{graphicx}
\usepackage{textcomp}
\usepackage{url}
\usepackage{float}
\usepackage{color}
\usepackage{cases}
\usepackage[normalem]{ulem}

\usepackage[titletoc,title]{appendix}

\usepackage[colorlinks=true, urlcolor=blue, anchorcolor=blue, citecolor=blue,filecolor=blue,linkcolor=blue,menucolor=blue
]{hyperref}

\usepackage{color}

\newcommand{\be}{\begin{equation}}
\newcommand{\ee}{\end{equation}}
\newcommand{\bea}{\begin{eqnarray}}
\newcommand{\eea}{\end{eqnarray}}

\usepackage{fixmath}

\begin{document}

\title{Path integrals for fractional Brownian motion and fractional Gaussian noise}

\author{Baruch Meerson}
\email{meerson@mail.huji.ac.il}
\affiliation{Racah Institute of Physics, Hebrew University of
Jerusalem, Jerusalem 91904, Israel}
\author{Olivier B\'{e}nichou}
\email{olivier.benichou@sorbonne-universite.fr}
\affiliation{Sorbonne Universit\'{e}, CNRS, Laboratoire de Physique Th\'{e}orique de la Mati\`{e}re Condens\'{e}e (UMR CNRS 7600), 4 Place Jussieu, 75252
Paris Cedex 05, France}
\author{Gleb Oshanin}
\email{gleb.oshanin@sorbonne-universite.fr}
\affiliation{Sorbonne Universit\'{e}, CNRS, Laboratoire de Physique Th\'{e}orique de la Mati\`{e}re Condens\'{e}e (UMR CNRS 7600), 4 Place Jussieu, 75252
Paris Cedex 05, France}
\affiliation{Dipartimento di Scienze Matematiche, Politecnico di Torino, Corso Duca degli Abruzzi 24, 10129 Torino, Italy}

\begin{abstract}
The Wiener's path integral plays a central role in the studies of Brownian motion.
Here we derive exact path-integral representations
for the more general \emph{fractional} Brownian motion (fBm) and for its time derivative process -- the fractional Gaussian noise (fGn). These  paradigmatic non-Markovian stochastic processes,  introduced by Kolmogorov, Mandelbrot and van Ness,
found numerous applications across the disciplines, ranging from anomalous diffusion in cellular environments to mathematical finance. Still, their exact path-integral representations were previously unknown. Our formalism exploits the Gaussianity of the fBm and fGn, relies on theory of singular integral equations and overcomes some technical difficulties by representing the action functional for the fBm in terms of the fGn for the sub-diffusive fBm, and in terms of the derivative of the fGn for the super-diffusive fBm. We also extend the formalism to include external forcing. The exact and explicit path-integral representations open new inroads into the studies of the fBm and fGn.

\end{abstract}
\maketitle

\textit{Introduction.} --
The importance of path integrals in theoretical physics is broadly recognized. Their application proved to be rewarding not only as a computational tool, both analytical and numerical, but also as a powerful and versatile conceptual framework. The notion of path integrals was introduced in the 1920s by Wiener \cite{Wiener} for  the Brownian motion (Bm). Since then  it helped uncover many nontrivial statistical properties of the Bm \cite{kac,satya,greg,vin,dean}.
Feynman reinvented path integrals in the 1940s  within his reformulation of quantum mechanics \cite{Feynman,hibbs,Feynman2}. He is also credited for making path integrals an intrinsic part of physicist's toolbox \cite{book0,wiegel,book1,zinn,book3,kamenevbook,book4,leticia}.

Path-integral representations of stochastic processes and fields
are especially useful in the studies of large-deviation statistics of physical quantities.
Performing a saddle-point evaluation of the pertinent path integral (which relies on a problem-specific
large parameter), one can determine the optimal, that is the most likely,  history of the system which dominates the statistics in question. This method of large deviation analysis
appears in different areas of physics under different names: the optimal fluctuation method, the instanton method, the weak-noise theory, the
macroscopic fluctuation theory, the dissipative WKB approximation, \textit{etc}. A full list of references on
different applications of this method would exceed a hundred.

\begin{figure}                                                                                    \includegraphics[width=0.15\textwidth,clip=]{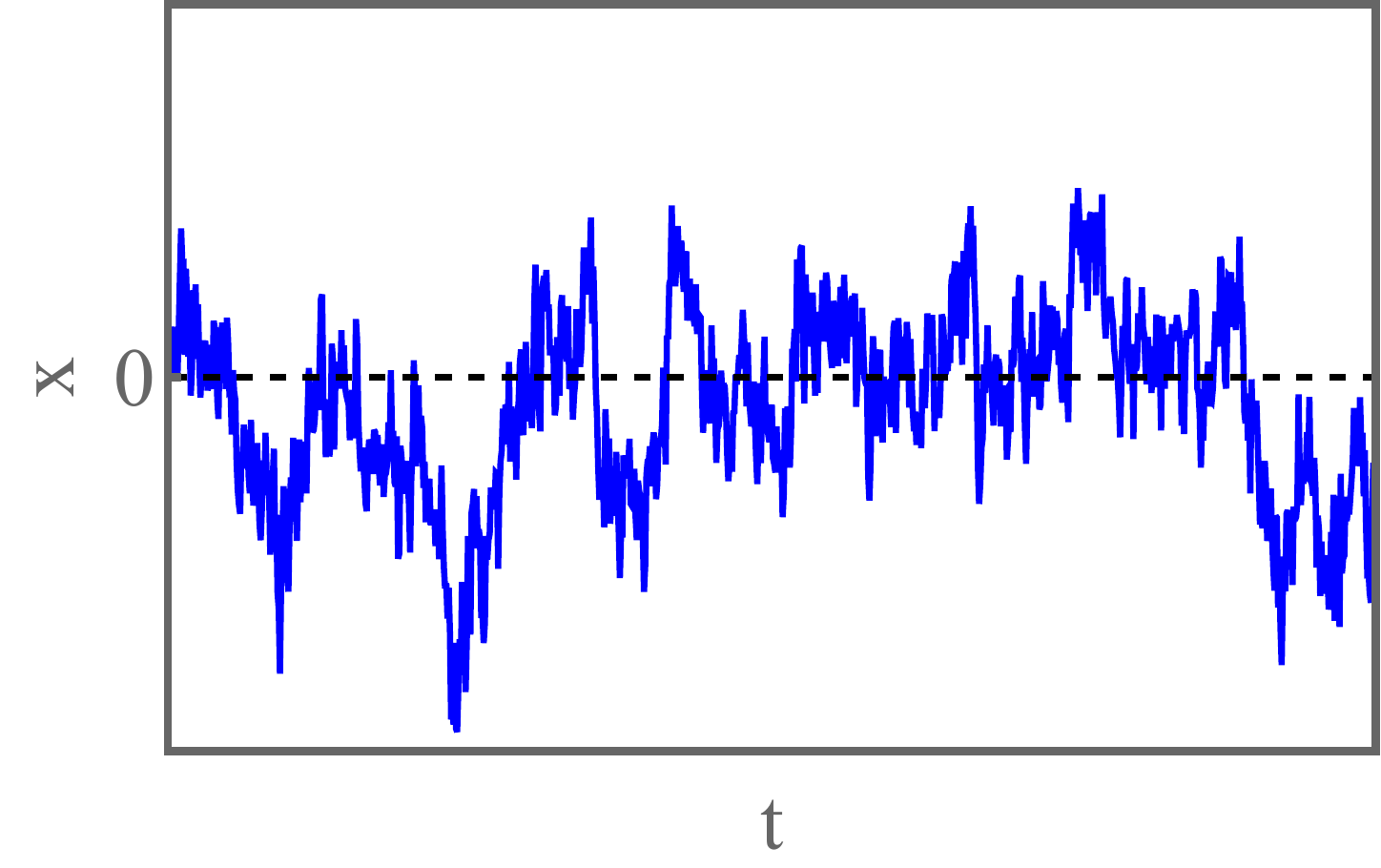}
\includegraphics[width=0.15\textwidth,clip=]{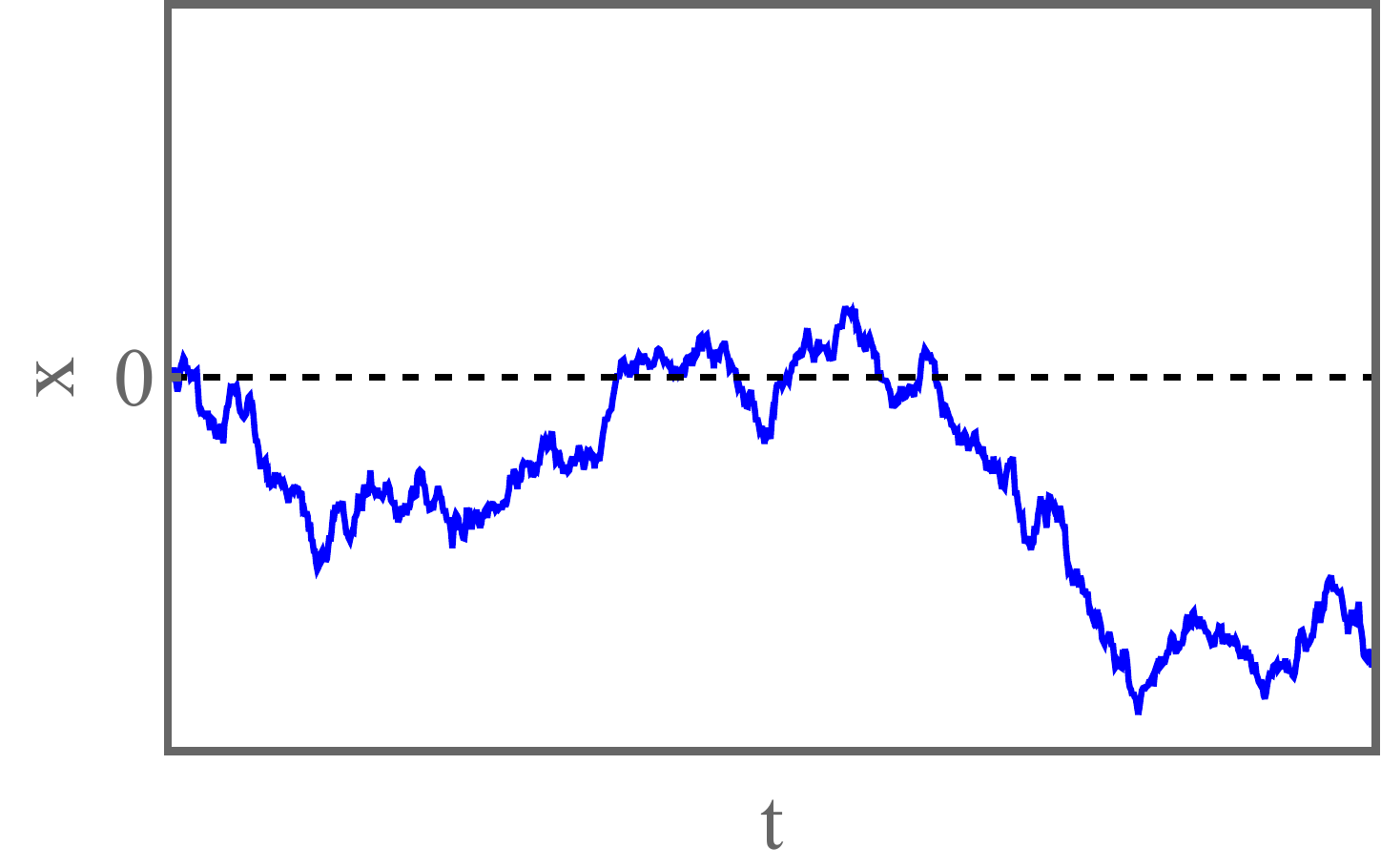}
\includegraphics[width=0.15\textwidth,clip=]{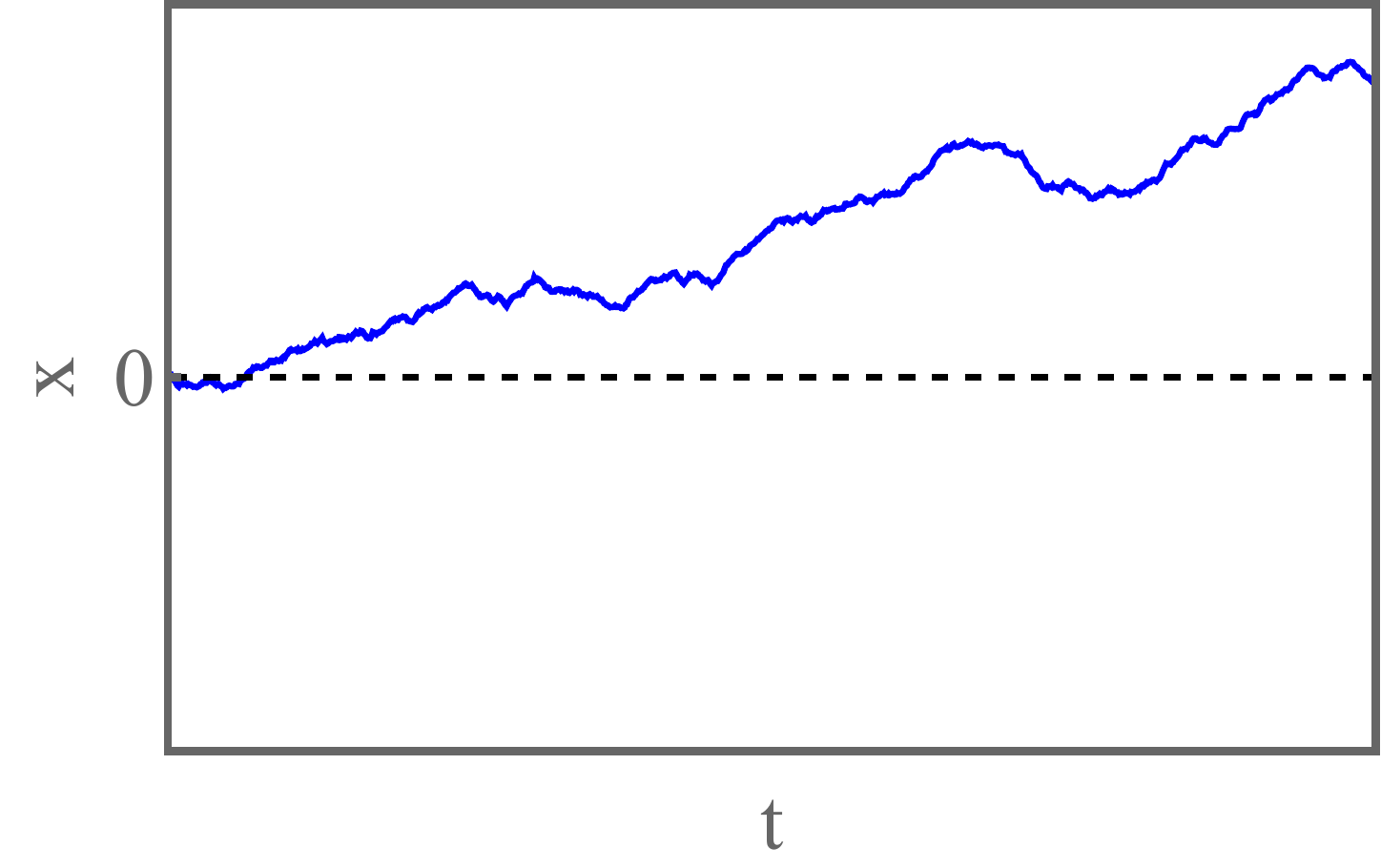}
\caption{Stochastic realizations of one-sided MvN fBm $x(t)$ for $H=1/4$ (top left) and $H=3/4$ (bottom). $H=1/2$ (top right) corresponds to the standard Bm.}
\label{realizations}
\end{figure}

The key object of a path-integral representation of Bm and its functionals is the probability density $P[x(t)]$ of a given realization of a Brownian trajectory $x(t)$, $P[x(t)] \sim \exp(-S[x(t)])$, where the action functional $S[x(t)]$ is given by the Wiener's formula \cite{Wiener}
\begin{equation}\label{Wiener}
S[x(t)]=\frac{1}{2}\int dt  \,\dot{x}^2(t)
\end{equation}
(the dot here and henceforth denotes the time derivative, and we set the diffusion coefficient to  $1/2$ for brevity). The local-in-time Wiener's action \eqref{Wiener} reflects the  Markovian nature of the Bm. The last two decades have witnessed a great
interest in the \emph{fractional} Brownian motion (fBm), introduced by Mandelbrot and van Ness \cite{Mandelbrot}, and earlier by Kolmogorov \cite{Kolmogorov}. The Mandelbrot-van Ness (MvN) fBm is a non-Markovian generalization of the Brownian motion which keeps the important properties of Gaussianity, stationarity of the increment, and dynamical scale invariance.  For the two-sided (that is, pre-thermalized) fBm, time $t$ is defined on the entire axis $|t|<\infty$. For the one-sided fBm $0 \leq t < \infty$.  Here the process starts at $t=0$, and there is no past.
Both versions of the fBm are zero-mean Gaussian processes (for convenience we set $x(0)=0$), and they are completely defined by their covariance functions
\begin{equation}\label{kappa}
\begin{split}
\!\kappa_2(t,t')\!=\!\langle x(t) x(t')\rangle\!&=\!\frac{1}{2}\left(|t|^{2H}\!+\!|t'|^{2H}\!-\!|t-t'|^{2H}\right) \,, \\
\kappa_1(t,t')\!=\!\langle x(t) x(t')\rangle\!&=
\!\frac{1}{2}\left(t^{2H}+t'^{2H}-|t-t'|^{2H}\right) \,.
\end{split}
\end{equation}
Here the subscript 1 and 2 stand for the one- and two-sided processes, respectively,  the angle brackets denote ensemble averaging, and $0<H<1$ is the Hurst index which quantifies the dynamical scale-invariance of the process \cite{Stanley} and its ruggedness. For $H < 1/2$ the fBm is sub-diffusive, \textit{i.e.} the mean-squared displacement  $\!\langle x^2(t) \rangle\! = t^{2H}$ grows sub-linearly with time. For $H > 1/2$ the fBm is super-diffusive. In the borderline case $H = 1/2$ one recovers the standard Bm. Figure \ref{realizations} presents examples of numerical stochastic realizations of fBm for $H=1/4$, $1/2$ (standard Bm)
and $3/4$.

The fractional Gaussian noise (fGn) was introduced by Mandelbrot and van Ness \cite{Mandelbrot} as the time-derivative of $x(t)$. That is, by definition, the fBm $x(t)$ obeys the Langevin equation $\dot{x}(t) = y(t)$, where $y(t)$ is the fGn. For $H < 1/2$ the fGn is anti-persistent (that is, it has negative autocorrelations).  For  $H > 1/2$ it is positively correlated. For $H=1/2$ the delta-correlated white Gaussian noise is recovered. The subsequent analysis covers both sub- and super-diffusive cases.

Multiple physical processes have been successfully modeled as fBm. These include fluctuating interfaces \cite{Krug}
dynamics in crowded fluids \cite{weiss,weiss2}, sub-diffusive dynamics of bacterial loci in a cytoplasm \cite{weber},
telomere diffusion in
the cell nucleus \cite{garini,prx},
modeling of conformations of serotonergic axons \cite{vojta},
diffusion of a tagged bead of a polymer \cite{Walter,Amitai}, translocation of a polymer through a pore \cite{Amitai,Zoia,Dubbeldam,Palyulin}, single-file
diffusion in ion channels \cite{Kukla,Wei,chanel}, \textit{etc.} A review can be found in \cite{ralf}.
In its turn, the fGn \cite{Mandelbrot} is used to model anti-persistent or persistent dependency structures in observed time series in many applications including hydrology \cite{Molz}, information theory \cite{Barton}, climate data analysis \cite{climate} and physiology \cite{physiol}, to mention a few.

By now the MvN fBm has become a standard model of anomalous diffusion in systems with memory. Still,  a satisfactory
path-integral representation of this process  is unavailable  \cite{without}.  This is in spite of the fact that, for non-Markovian but Gaussian processes, such as the MvN fBm, there is a straightforward path  \cite{zinn}  to constructing an analog of Eq. \eqref{Wiener}. It involves the determination of a (highly singular)  nonlocal kernel, inverse to the covariance function \eqref{kappa}, via solving a singular integral equation [such as Eq. \eqref{int} below]. For MvN fBm this equation is hard to deal with analytically, which explains the scarcity of results on path-integral representations of the fBm \cite{RLfBm}.

These technical obstacles were circumvented in the early work \cite{bead}, where a nonlocal analog of the Wiener's action \eqref{Wiener} was derived, by a different method, in the particular case of the dynamics of a tagged bead in an infinitely long pre-thermalized Rouse polymer \cite{bead}. Under some natural assumptions this non-Markovian system is equivalent to a fluctuating interface in one dimension, and the latter is known to be  describable by the MvN fBm with the Hurst exponent $H=1/4$ \cite{Krug}. The action, calculated in Ref. \cite{bead}, is given, up to a constant factor, by the expression
 \begin{equation}
 \label{bead}
 S_{\text{bead}}[x(t)] \sim \iint  \frac{dt_1 dt_2}{|t_1 - t_2|^{1/2}} \, \dot{x}(t_1)  \dot{x}(t_2)\,,
 \end{equation}
see also Ref. \cite{nech}. We should also mention a series of works \cite{19,20,21,22,23,khadem} aimed at determining $S[x(t)]$ for the one-sided MvN fBm  in the form of a perturbation expansion around the Wiener's action \eqref{Wiener}. By construction, such an expansion, based on the small parameter $|H-1/2|\ll 1$,  is quite limited in its validity.

In this work we find exact and explicit non-local analogs of the Wiener's action \eqref{Wiener} for the  MvN
fBm: for arbitrary $0<H<1$ and for both two-sided and one-sided versions of the fBm. We also extend the path integrals to include overdamped motion of the particle under external force. We achieve these goals by seeking, from the start, the action functional for the fBm in terms of its time derivative processes: the first-derivative process (that is, the   fGn) for the sub-diffusive fBM, and the second-derivative process for the super-diffusive fBm \cite{motivation}. Our formalism fully exploits the Gaussianity of the fBm and relies on the well-established theory of singular integral equations, see \textit{e.g.} Refs. \cite{kilbas,polyanin}. The resulting path integrals are convenient to work with, as they involve only mildly singular kernels.  Finally, we extend the formalism to include external forcing.

\textit{General expressions and main results.} -- Quite generally, the action functional $S[X(t)]$ of a Gaussian process $X(t)$ on a time interval $\Omega$ can be represented as \cite{zinn}
\begin{equation}\label{actiongen}
S[X(t)] = \frac{1}{2}  \int_{\Omega} dt_1  \int_{\Omega} dt_2 \, \mathcal{K}(t_1,t_2) \,X(t_1) \,X(t_2) \,.
\end{equation}
The kernel $\mathcal{K}(t_1,t_2)$ (a symmetric function of $t_1$ and $t_2$) is the inverse
of the covariance function $\kappa(t_1,t_2)$ of the process $X(t)$:
\begin{equation}
\label{int}
\int_{\Omega} dt_1 \, \kappa\left(t_1,t_3\right) \, \mathcal{K}(t_1,t_2)  = \delta\left(t_2 - t_3\right) \,.
\end{equation}
Once  $\mathcal{K}(t_1,t_2)$  is known, the action functional \eqref{actiongen} is completely defined, giving the probability density $P[X(t)]$ of a given realization of the process $X(t)$. Now we present our main results for the action functionals of the MvN fBm $x(t)$. They have different forms for the sub-diffusive and super-diffusive fBm, and for the two- and one-sided processes.

We start with the sub-diffusion. For the two-sided sub-diffusive ($0<H<1/2$) fBm $x(t)$,
the action $S = S[x(t)]$ is given by
\begin{equation}
\label{S2sub}
S = \frac{{\rm cot}(\pi H)}{4 \pi H} \int^{\infty}_{-\infty}
\int^{\infty}_{-\infty} \frac{dt_1 dt_2}{|t_1 - t_2|^{2H}} \,
\dot{x}(t_1) \dot{x}(t_2)\,.
\end{equation}
For the one-sided sub-diffusive fBm we obtain
\begin{equation}
\label{S1sub}
S \!=\! \frac{{\rm cot}(\pi H)}{4 \pi H} \!\int^{\infty}_{0}    \int^{\infty}_{0} \!\!  \frac{ dt_1 dt_2\,I_{z}\left(\frac{1}{2}-H,H\right)}{|t_1 - t_2|^{2H}} \dot{x}(t_1) \dot{x}(t_2) ,
\end{equation}
where  $I_z(a,b)$ is the regularized incomplete beta function
\begin{equation}
\label{betareginc}
I_z(a,b) = \frac{\Gamma(a + b)}{\Gamma(a) \Gamma(b)} \int^z_0 x^{a - 1} (1- x)^{b-1} dx \,,
\end{equation}
$\Gamma(\dots)$ is the gamma function, and $z=4 t_1 t_2(t_1 + t_2)^{-2}$.

As one can see, the action functionals \eqref{S2sub} and \eqref{S1sub} are non-local in time and written in terms of the fGn $\dot{x}(t)$, rather than in terms of $x(t)$ itself. The expression \eqref{S1sub} for the one-sided case is more complicated, than that for the two-sided one, Eq. \eqref{S2sub}. In particular, the two-sided kernel in Eq. \eqref{S2sub} is a difference kernel, which reflects the stationarity in time of the two-sided derivative process, the fGn. The one-sided kernel \eqref{S1sub} is not a difference kernel in spite of the stationarity of the fGn. The non-stationarity, however, is temporary, as it is caused by a transient created by
the initial condition $x(t=0)=0$. Indeed, in the limit of $t_1, t_2 \to \infty$, and $t_1-t_2 =\text{const}$, $z$ tends to $1$, the one-sided kernel coincides with the two-sided one, and the stationarity is restored.

In the limiting case $H=1/2$, the kernels in Eqs. \eqref{S2sub} and \eqref{S1sub} become delta-functions and yield the classical Wiener's formula \eqref{Wiener}, as we show in Ref. \cite{SM}.

For $H = 1/4$ Eq. \eqref{S2sub} has the same functional form as the two-sided expression \eqref{bead}, as to be expected in view of the pre-thermalization of the Rouse polymer \cite{bead}. We also remark that  Eq. \eqref{S2sub} was postulated in Ref. \cite{nech} as an effective Hamiltonian of topologically
stabilized polymers in melts, permitting to cover various conformations ranging from
ideal Gaussian coils to crumpled globules. Our derivation validates their approach.

Now we present our results for the super-diffusive fBm, $1/2<H<1$. In the two-sided case we obtain
\begin{equation}
\label{S2super}
S \!=\! \frac{\sigma(H)}{2}\, \!\int^{\infty}_{-\infty}
\! \int^{\infty}_{-\infty} dt_1\, dt_2\, |t_1 - t_2|^{2-2H}\, \ddot{x}(t_1) \, \ddot{x} (t_2)\,,
\end{equation}
where
\begin{equation}\label{sigmaH}
\sigma(H)=-\frac{{\rm cot}(\pi H)}{4 \pi H(1-H) (2H-1)} \,,
\end{equation}
a positive function. For  the one-sided case
\begin{eqnarray}
  S &=&  \frac{\sigma(H)}{2}\,  \int^{\infty}_0  \int^{\infty}_0  dt_1 \, dt_2  \, |t_1 - t_2|^{2-2H}   \nonumber\\
  &\times &  I_{z'} \left(\frac{3}{2}-H,2H-2\right) \ddot{x}(t_1) \ddot{x}(t_2) \,,
  \label{S1super}
\end{eqnarray}
where $ z'  = {\rm min}(t_1,t_2)/{\rm max}(t_1,t_2)$. Again, the expressions in Eqs. \eqref{S2super} and \eqref{S1super} are non-local in time, but now they are written in terms of $\ddot{x}(t)$, that is in terms of the first derivative of the fGn. The two-sided kernel is a difference kernel.  The one-sided kernel is not, but it approaches the difference form following an initial transient. Also, the classical Wiener's form \eqref{Wiener} is recovered in the limit $H \to 1/2$ \cite{SM}.

Expressions \eqref{S2sub}-\eqref{S1super}, alongside with Eqs.  \eqref{potentialsub}  and  \eqref{actionforcesuper}  below, represent the main results of this work.   Here we present derivations of Eqs. \eqref{S2sub} and \eqref{S2super} for the two-sided sub-diffusive and super-diffusive fBm, respectively. The derivation of the (a bit more bulky) one-sided expressions in Eqs. \eqref{S1sub} and \eqref{S1super} is relegated to the SM \cite{SM}.

\textit{Sub-diffusion} -- Here we work directly with the fGn. Its covariance function $c(t_1,t_2)$ can be readily calculated:
\begin{eqnarray}
&&c(t_1,t_2)=\langle y(t_1) y(t_2)\rangle =\langle \dot{x}(t_1) \dot{x}(t_2)\rangle \nonumber \\
  &&= \frac{\partial^2}{\partial t_1 \partial t_2} \langle x(t_1) x(t_2)\rangle  =  \frac{d}{d\tau}\left( H |\tau|^{2H-1}
  \text{sgn}\,\tau\right)\,,\label{corry}
\end{eqnarray}
where $\tau = t_1-t_2$, and we used Eq.~\eqref{kappa}. Equation \eqref{corry} holds both for the two-sided and the one-sided process and for all $0<H<1$. Notably, the fGn is a stationary process.
For $H=1/2$ Eq.~\eqref{corry} gives $c(\tau) =(1/2) (d/d\tau) \,\text{sgn}\,\tau=\delta(\tau)$, as to be expected for the white noise.

Let us denote by $\mathcal{C}(\tau)$ the kernel inverse to $c(\tau)$. For the two-sided process, $\mathcal{C}(\tau)$ is defined by the equation $\int_{-\infty}^{\infty} d\tau   \, c\left(\tau-t\right) \, \mathcal{C}(\tau) = \delta (t)$
or, in the explicit form,
\begin{equation}\label{inteqn1}
\int_{-\infty}^{\infty} d \tau \,\mathcal{C}(\tau) \frac{d}{d\tau} \left[|\tau-t|^{2H-1} \text{sgn} (\tau-t)\right] =\frac{1}{H}\,\delta(t).
\end{equation}
Integrating by part and assuming that the boundary terms are zero (as can be verified \textit{a posteriori}), we arrive at the integral equation
\begin{equation}\label{inteqn2}
\int_{-\infty}^{\infty} d \tau \,\frac{\text{sgn} (\tau-t)}{|\tau-t|^{1-2H}} \,\mathcal{D}(\tau) =-\frac{1}{H}\,\delta(t)
\end{equation}
for the unknown function $\mathcal{D}(\tau) = d \mathcal{C}(\tau)/d\tau$.
The solution can be found in Ref. \cite{kilbas}:
\begin{equation}\label{solD2}
\mathcal{D}(\tau) =\frac{d\mathcal{C}(\tau)}{d\tau} = \frac{\text{cot}\, (\pi H)}{2\pi H} \frac{d}{d\tau}\frac{1}{|\tau|^{2H}}\,.
\end{equation}
Getting rid of the $\tau$-derivative and using the fact that the kernel must vanish at $|\tau| \to \infty$, we obtain
\begin{equation}\label{solC2}
\mathcal{C}(\tau) = \frac{\text{cot}\, (\pi H)}{2\pi H} \frac{1}{|\tau|^{2H}}\,.
\end{equation}
The ensuing Gaussian action functional \eqref{actiongen}, written in terms of $X(t)=\dot{x}(t)$,
yields the announced equation~\eqref{S2sub}.

\textit{Super-diffusion} --  Here we work with the second-derivative process $z(t) = \ddot{x}(t)$. Its covariance is
\begin{equation}
q(t_1,t_2)=\frac{d^3}{d\tau^3}\left[ H |\tau|^{2H-1} \text{sgn}(\tau)\right]\,.\label{q}
\end{equation}
For the two-sided process  the inverse kernel $\mathcal{Q}(t_1,t_2)$ is defied by the equation $\int_{-\infty}^{\infty} d\tau   \, q\left(\tau-t\right) \, \mathcal{Q}(\tau)= \delta (t)$,
or, in the explicit form,
\begin{equation}\label{inteqn10}
\int_{-\infty}^{\infty} d \tau \,\mathcal{Q}(\tau) \frac{d^3}{d\tau^3} \left[|\tau-t|^{2H-1} \text{sgn} (\tau-t)\right] =\frac{1}{H}\,\delta(t).
\end{equation}
Integrating three times by part and assuming that the boundary terms are zero (as verified \textit{a posteriori}), we arrive at the equation
\begin{equation}\label{inteqn11}
\int_{-\infty}^{\infty} d \tau \,\frac{\text{sgn} (\tau-t)}{|\tau-t|^{1-2H}} \,\mathcal{Z}(\tau) =-\frac{1}{H}\,\delta(t),
\end{equation}
where $\mathcal{Z}(\tau) = d^3 \mathcal{Q}(\tau)/d\tau^3$. This is exactly the same equation as Eq. \eqref{inteqn2}, but now $1-2H<0$. It is convenient to rewrite this equation as
\begin{eqnarray}
 -\int_{-\infty}^{t} d \tau \,\left(t-\tau\right)^{2H-1} \mathcal{Z}(\tau)
 &+& \int_{t}^{\infty} d \tau \,\left(\tau-t\right)^{2H-1} \mathcal{Z}(\tau)
 \nonumber \\ &=& -\frac{1}{H}\,\delta(t)  \label{inteqn3}
\end{eqnarray}
and differentiate both sides of Eq.~(\ref{inteqn3}) with respect to $t$. The resulting equation,
\begin{equation}\label{inteqn4}
\int_{-\infty}^{\infty} d \tau \,\frac{\mathcal{Z}(\tau)}{|\tau-t|^{2-2H}}  =\frac{1}{H(2H-1)}\,\delta^{\prime}(t),
\end{equation}
is solvable \cite{kilbas}, and we obtain
\begin{equation}\label{sol10}
\mathcal{Z}(\tau)= -\frac{\text{cot}(\pi H)}{2\pi H (2H-1)}\, \frac{d^2}{d\tau^2}\frac{\text{sgn}\,\tau}{|\tau|^{2H-1}}\,. \end{equation}
Integrating this expression over $\tau$ three times and using account the fact that the kernel must vanish at $|\tau| \to \infty$,
we obtain the desired inverse kernel:
\begin{equation}\label{kernelmore}
\mathcal{Q}(\tau) =    \sigma(H)\,|\tau|^{2(1-H)}\,,
\end{equation}
where $\sigma(H)$ is defined in Eq.~\eqref{sigmaH}. The resulting Gaussian action functional \eqref{actiongen}, written in terms of $X(t)=z(t)\equiv\ddot{x}(t)$,
yields the announced Eq.~\eqref{S2super}.

\textit{External force}  -- An important extension of this formalism deals with situations where the fBm of a
particle is accompanied by
its overdamped motion under external force $f(x)$. A natural approach to modelling this situation
employs the non-Markovian Langevin
equation \cite{Metzler2021}
\begin{equation}\label{Langevin}
\dot{x}(t) = f[x(t)] + y(t)\,,
\end{equation}
where the noise term $y(t)$ describes fGn.  When the external
force $f(x)$ is confining, the $x$-distribution approaches a steady state. This steady state, however, is non-Boltzmann. Therefore, not surprisingly, it violates the fluctuation-dissipation theorem \cite{Metzler2021}.
As the fGn $y(t)$ is a Gaussian process, a natural path-integral representation for Eq.~\eqref{Langevin} is provided by
the action functional
\begin{eqnarray}
  S[x(t)] &=& \frac{1}{2} \int_{-\infty}^{\infty} dt_1 \int_{-\infty}^{\infty} dt_2 \,
  \mathcal{C}(t_1-t_2)\,\{\dot{x}(t_1)-f[x(t_1)]\} \nonumber \\
 &\times&   \{\dot{x}(t_2)-f[x(t_2)]\} \,,
\label{potentialsub}
\end{eqnarray}
where $\mathcal{C}(\tau)$ is the inverse kernel for the fGn, given by Eq.~\eqref{solC2}. Here we assumed a two-sided sub-diffusive fBm.

For a super-diffusive fBm a suitable non-Markovian Langevin equation can be obtained by a formal differentiation of Eq. \eqref{Langevin}
with respect to time, leading to
\begin{equation}\label{Langevinsuper}
\ddot{x}(t) = f'[x(t)]\dot{x}(t) + z(t)\,,
\end{equation}
where $f'(x)\equiv df(x)/dx$, and the noise term $z(t)$ is the time derivative of the fGn.  The corresponding path integral
for the two-sided process is given by the action functional
\begin{eqnarray}
&&S[x(t)] = \frac{1}{2} \int_{-\infty}^{\infty} dt_1 \int_{-\infty}^{\infty} dt_2\,\mathcal{Q}(t_1-t_2)  \nonumber \\
&\times&\!\!\left[\ddot{x}(t_1)-f'[x(t_1)]\,\dot{x}(t_1)\right] \left[\ddot{x}(t_2)-f'[x(t_2)]\,\dot{x}(t_2)\right], \label{actionforcesuper}
\end{eqnarray}
where $\mathcal{Q}(\tau)$ is given by Eq. \eqref{kernelmore}. Expressions similar to Eqs. \eqref{potentialsub} and \eqref{actionforcesuper}, but with the one-side kernels as in Eqs. \eqref{S1sub} and \eqref{S1super}, hold for the one-sided sub- and super-diffusive fBm, respectively.

\textit{Summary.} --  We generalized the classical Wiener's path integral for the Bm and
found exact path-integral representations for the two-sided and one-sided MvN fBm for the whole range $0<H<1$ of the Hurst exponent. We also extended the formalism to include external forcing. The exact and explicit path-integral representations open new inroads into analytical and numerical studies of fBm -- an important paradigm of scale-invariant stochastic processes with memory -- in a multitude of applications in natural sciences, technology and finance.

\textit{Acknowledgments.} -- We are grateful to P. Chigansky, D. S. Dean, S. N. Majumdar and K. L. Sebastian for useful discussions. B. M. was supported by the Israel Science Foundation (Grant No. 1499/20).

\appendix

\begin{widetext}

\section{Supplemental Material}
\label{SM}

\noindent Here we present some details of derivations of the results obtained for the one-sided case
 and also show that in the limit $H \to 1/2$ the expressions (6), (7), (9) and (11) converge to the Wiener result in Eq. (1).


\renewcommand{\theequation}{S\arabic{equation}}
\setcounter{equation}{0}

\section{Two-sided fBm}

\subsection{Sub-diffusion, $0 < H < 1/2$. The limit of $H\to 1/2$}

\noindent
In order to take the limit $H\to 1/2$, we take advantage of the identity
\begin{align}
\frac{1}{|t_1 - t_2|^{2H}} = \frac{1}{(1-2H)} \frac{d}{dt_1} \frac{{\rm sgn}(t_1 - t_2)}{|t_1 - t_2|^{2H-1}} \,,
\end{align}
which permits us to formally rewrite the kernel in Eq. (3) as
\begin{align}
\frac{{\rm cot}(\pi H)}{2 \pi H} \frac{1}{|t_1 - t_2|^{2H}} =  \frac{{\rm cot}(\pi H)}{2 \pi H(1-2H)} \frac{d}{dt_1} \frac{{\rm sgn}(t_1 - t_2)}{|t_1 - t_2|^{2H-1}} \,.
\end{align}
Taking the limit $H \to 1/2$ in the both sides of the latter equality
and noticing that
\begin{align}
\lim_{H \to 1/2} \frac{{\rm cot}(\pi H)}{(1- 2H)} = \frac{\pi}{2} \,,
\end{align}
we get
\begin{align}
\lim_{H \to 1/2} \frac{{\rm cot}(\pi H)}{2 \pi H} \frac{1}{|t_1 - t_2|^{2H}} &= \lim_{H \to 1/2}  \frac{{\rm cot}(\pi H)}{2 \pi H(1-2H)} \frac{d}{dt_1} \frac{{\rm sgn}(t_1 - t_2)}{|t_1 - t_2|^{2H-1}} \\
&=  \frac{1}{2} \frac{d}{dt_1} {\rm sgn}(t_1 - t_2) =  \delta(t_1 - t_2) \,,
\end{align}
which yields the Wiener expression (1).

\subsection{Super-diffusion, $1/2 < H < 1$. The limit of $H\to 1/2$}

We turn to the limit $H \to 1/2$ directly in Eq. (22) in the main text to get
\begin{align}
\lim_{H \to 1/2}\mathcal{Z}(\tau)= \mathcal{Z}_{H=1/2}(\tau) = \frac{1}{2} \frac{d^2}{d \tau^2} {\rm sgn}(\tau) = \frac{d}{d \tau} \delta(\tau)
\end{align}
The action $S$ written in terms of the derivative of the fractional Gaussian noise involves the kernel function $\mathcal{Q}$, which is given by a triple integral of $\mathcal{Z}_{H=1/2}(\tau)$ [see Eq. (23)].   Consequently, the action has the form
\begin{align}
S=\frac{1}{2} \int^{\infty}_{-\infty} \int^{\infty}_{-\infty} dt_1 dt_2 \ddot{x}(t_1) \ddot{x}(t_2) \int^{t_1}_{-\infty} \int^{t_2}_{-\infty} d\tau_1 d\tau_2 \delta(\tau_1 - \tau_2) \,.
\end{align}
Integrating the latter expression by parts, we arrive at the Wiener's result in Eq. (1).

\section{One-sided fBm}

\noindent Here we present brief derivations of our expressions (7) and (11) of the main text.

\subsection{Sub-diffusion, $0 < H < 1/2$}
For the one-sided sub-diffusive fBm the inverse kernel $\mathcal{C}(t_1,t_2)$ is not a difference kernel, and Eq.~(13) gives way to the equation
\begin{eqnarray}
\int_{0}^{\infty} d t_1 \,\mathcal{C}(t_1,t_2) \frac{d}{dt_1} \left[|t_1-t_3|^{2H-1} \text{sgn}
(t_1-t_3)\right]  =  \frac{1}{H}\,\delta(t_2-t_3)\,.&&
\end{eqnarray}
Integrating by part, we arrive at
\begin{equation}\label{inteqn2a}
\int_{0}^{\infty} d t_1 \,\frac{\text{sgn} (t_1-t_3)}{|t_1-t_3|^{1-2H}} \,\mathcal{D}(t_1,t_2)
=-\frac{1}{H}\,\delta(t_2-t_3),
\end{equation}
where $\mathcal{D}(t_1,t_2) = (\partial/\partial t_1) \,\mathcal{C}(t_1,t_2)$.
The solution can be found in Ref. [47]. Getting rid of the $t_1$-derivative, we
obtain $\mathcal{C}(t_1,t_2)$ in the following three alternative (but equivalent) forms
\begin{align}\label{q1}
\mathcal{C}(t_1,t_2) &= \frac{(t_1 t_2)^{-H} \sqrt{z'}}{H(1-2H) \sin(\pi H) \Gamma(2H)
\Gamma^2\left(\frac{1}{2}-H\right)} \,_2F_1\left(1,H+\frac{1}{2};\frac{3}{2}-H;z'\right) \,, \quad z' = \frac{{\rm min}(t_1,t_2)}{{\rm max}(t_1,t_2)} \\
&= \frac{{\rm cot}(\pi H) (t_1 t_2)^{-H} \sqrt{z}}{2 \pi H (1-2H) B(1/2-H,2H)} \,_2F_1\left(\frac{1}{2},1;\frac{3}{2}-H;z\right) \,, \quad z = \frac{4 t_1 t_2}{(t_1 + t_2)^2} \\
& = \frac{{\rm cot}(\pi H)}{2 \sqrt{\pi} \Gamma(1/2-H) \Gamma(1+H)} \frac{B_z(1/2-H,H)}{|t_1 - t_2|^{2H}} \,,
\end{align}
where $B(a,b)$ and $B_z(a,b)$ are the complete and incomplete beta-functions, respectively, and
$_2F_1(\dots)$ is the hypergeometric function.
Recalling next the definition of the regularized incomplete beta-function [see Eq. (8)], we obtain our result in Eq. (7).

\vspace{0.25 cm}
\textbf{Limit $H \to 1/2$.}  To take the limit $H \to 1/2$ in Eq. (7) in the main text it is expedient to use
the representations of the kernel $\mathcal{C}(t_1,t_2)$ given in the second line in Eq. (\ref{q1}). We formally rewrite the Gauss hypergeometric function entering this representation as
\begin{align}
\,_2F_1\left(\frac{1}{2},1;\frac{3}{2}-H;z\right) = \frac{\Gamma(H) \Gamma(3/2-H)}{\sqrt{\pi} (1-z)^{H} z^{1/2-H}} - \frac{(1/2-H)}{H} \,_2F_1\left(\frac{1}{2},1;1+H;1 - z\right)
\end{align}
such that,  after some algebra, the kernel $\mathcal{C}(t_1,t_2) $ can be cast into the form
\begin{align}
\mathcal{C}(t_1,t_2)  = \frac{{\rm cot}(\pi H)}{2 \pi H} \frac{1}{|t_1 - t_2|^{2 H}} - \frac{{\rm cot}(\pi H)}{2 \pi H^2 B(1/2-H,2H)} \frac{(t_1 t_2)^{1/2-H}}{(t_1 + t_2)} \,_2F_1\left(\frac{1}{2},1;1+H;\left(\frac{t_1-t_2}{t_1+t_2}\right)^2\right)
\end{align}
Further on, we observe that the numerical $H$-dependent amplitude in the second term in the latter expression vanishes in the limit $H \to 1/2$  as
\begin{align}
\frac{{\rm cot}(\pi H)}{2 \pi H^2 B(1/2-H,2H)} \simeq 2 \left(\frac{1}{2} - H\right)^2 \,,
\end{align}
which signifies that this term does not contribute in this limit. On the contrary, as demonstrated above, the first term converges to the delta-function which ensures that the action in Eq. (7) in the main text converges to the Wiener's result in Eq. (1).

\subsection{Super-diffusion, $1/2 < H < 1$}

Here we present a derivation of Eq. (11) and also check the limit of $H\to 1/2$. In particular, this derivation
highlights the reason why the representation of the action in terms of $\ddot{x}$ is
advantageous.

We start the derivation by representing the action in terms of the first derivative of the fBm, that is in terms of the fGn:
\begin{align}
\label{q2}
S = \frac{1}{2} \int^{\infty}_0 \int^{\infty}_0 dt_1 dt_2 \mathcal{C}(t_1,t_2) \dot{x}(t_1) \dot{x}(t_2) \,.
\end{align}
The inverse kernel $\mathcal{C}(t_1,t_2)$ obeys the integral equation
\begin{align}
\label{q3}
\int^{\infty}_0 \frac{dt_1 \, \mathcal{C}(t_1,t_2)}{|t_1 - t_3|^{2-2H}} = \frac{\delta(t_2 - t_3)}{H (2H-1)} \,.
\end{align}
The explicit solution of this equation can be found in Ref. \cite{Lundgren}. After straightforward transformations, it reads:
\begin{align}
\label{q4}
\mathcal{C}(t_1,t_2) &= \frac{1}{2 H \sin(\pi H) \Gamma(2H) \Gamma^2(3/2-H)} \frac{d^2}{dt_1 dt_2} \int_0^{{\rm min}(t_1,t_2)} \frac{d\tau}{(t_1 - \tau)^{H-1/2} (t_2 - \tau)^{H-1/2}} \\
&= \frac{1}{2 H \sin(\pi H) \Gamma(2H) \Gamma^2(3/2-H)} \frac{d^2}{dt_1 dt_2}   |t_1 - t_2|^{2 - 2H} B_{z'}(3/2-H,2H-2) \,, \quad z'= \frac{{\rm min}(t_1,t_2)}{{\rm max}(t_1,t_2)} \,.
\end{align}
We observe that the function $\mathcal{C}(t_1,t_2)$, defined in the right-hand-side of Eq. (\ref{q4}), contains a non-integrable singularity at $t_1 = t_2$, which shows why the representation of the action in Eq. (\ref{q2}) in terms of the first derivatives $\dot{x}(t)$ is problematic. To get a regular result, we integrate Eq. (\ref{q2}) by part,
\textit{i.e.} express it in terms of the second derivatives $\ddot{x}(t_1)$ and $\ddot{x}(t_2)$. Then, using the definition of the regularized incomplete beta-function in Eq. (8), we arrive at the final result in Eq. (11).

\vspace{0.25 cm}
\textbf{Limit $H \to 1/2$.} The limit $H \to 1/2$ can be conveniently taken in the expression given in the first line of Eq. (\ref{q4}). This yields
\begin{align}
\lim_{H \to 1/2} \mathcal{C}(t_1,t_2) =  \frac{d^2}{dt_1 dt_2}  {\rm min}(t_1,t_2) = \delta(t_1 - t_2)  \,,
\end{align}
and we recover the Wiener's expression in Eq. (1).

\end{widetext}


\begin{thebibliography} {99}
\bibitem{without} In simple large-deviation problems, analyzed with the optimal fluctuation method, the explicit knowledge of the inverse kernel can be unnecessary. This happens when
the ensuing nonlocal Euler-Lagrange equation  (a linear integral equation) can be transformed into a form containing only the covariance function of the process \cite{baruch,longtimeLD,escape}.
For more complicated problems, however, no such transformation exists, and the knowledge of the inverse kernel is indispensable.
\bibitem{RLfBm} For the Riemann-Liouville fBm (a different mathematical model of anomalous diffusion, introduced by Levy \cite{Levy}), the action $S[x(t)]$ can be determined by employing fractional calculus, as was done in Refs. \cite{seb,san} and in Chapter 10 in Ref. \cite{book4}. Also, in Refs. \cite{Calvo,Wiopaper} a path-integral representation was found for the fractional Levy motion    \cite{Calvo}, still another mathematical model. All these results, however, are irrelevant to the MvN fBm studied in our work.
\bibitem{motivation} Indeed, although $x(t)$ is a Gaussian process, the Wiener's action \eqref{Wiener} is expressed through $\dot{x}(t)$, rather than through $x(t)$. Furthermore, one can rewrite Eq. \eqref{Wiener} as $S[x(t)]=(1/2)\iint  dt_1 dt_2 \, \delta(t_1-t_2) \dot{x}(t_1) \dot{x}(t_2)$, with a delta-function kernel.  In the $x$-representation the kernel, $\partial^2_{t_1t_2}\, \delta(t_1-t_2)$, is more singular and less convenient to work with.
\end{thebibliography}
\end{document}